\documentstyle[epsf,twocolumn,prl,aps]{revtex}
\begin{document}
\draft
\preprint{YITP-96-2}
\title{Exact solution and spectral flow for
twisted Haldane-Shastry model}
\author{ Takahiro Fukui \cite{JSPS}\cite{Ema}} 
\address{Yukawa Institute for Theoretical Physics, Kyoto University,
Kyoto 606-01, Japan}
\author{Norio Kawakami}
\address{Department of Applied Physics,
and Department of Material and Life Science,\\
Osaka University, Suita, Osaka 565, Japan}
\date{January 30, 1996}
\maketitle
\begin{abstract}
The exact solution of the spin chain model with $1/r^2$ 
exchange is found for twisted boundary conditions.
The spectrum thus obtained can be reproduced  by the
asymptotic Bethe ansatz. The spectral flow of each eigenstate 
is determined  exactly as a function of the twist
angle. We find that the period  $4\pi$ for the ground state 
nicely fits in with the notion of fractional exclusion statistics.
\end{abstract}


Quantum integrable systems with $1/r^2$
interaction \cite{ISMD1,ISMD2,Hal,Sha,Kura} 
have attracted much interest recently, providing
us with various interesting notions such as fractional 
exclusion statistics\cite{Hal_f}, Yangian symmetry
\cite{HHTBP}, etc.  In particular, it is quite interesting
to ask whether one can observe novel features inherent in fractional 
exclusion statistics  explicitly in physical quantities.
To examine this problem, the response to external gauge
fields, which is equivalent to the response to 
twisted boundary conditions \cite{Twi}, is expected to
play a vital role. This naturally motivates us to 
investigate quantum $1/r^2$ lattice models\cite{Hal,Sha,Kura}
with twisted boundary conditions.
In contrast to the models with short-range 
interactions, however, it is not trivial to study the 
quantum $1/r^2$ models with twisted boundary conditions
because the long-range nature of interaction makes
the problem rather complicated.

In this letter, we generalize the Haldane-Shastry spin model
with $1/r^2$ exchange\cite{Hal,Sha} to include twisted 
boundary conditions. We find the exact solution of this model
{\it for every rational value of the twist angle in unit of $2\pi$}.
The energy spectrum thus obtained can be 
precisely reproduced by the asymptotic Bethe ansatz.
We determine the spectral flow of the energy spectrum
as a function of the twist angle. In particular, it is found that
the period $4\pi$ deduced for the ground state 
nicely  fits in with the notion of fractional 
exclusion statistics\cite{Hal_f}.
This demonstrates that {\it one can
observe fractional exclusion statistics explicitly} in the 
spectral flow for the Haldane-Shastry model,
although such a property is  generally hidden by 
irrelevant perturbations in ordinary interacting models
\cite{SutSha,FukKaw}.


Let us introduce the model Hamiltonian describing
the spin chain with $1/r^2$ exchange\cite{Hal,Sha}
\begin{eqnarray}
H=&&\frac{1}{2}\sum_{n=1}^N\sum_{n'=1}^{N-1}\sum_{l=-\infty}^\infty
\frac{1}{(n'+lN)^2}\times \nonumber\\
&&\left(S_n^xS_{n+n'+lN}^x+S_n^yS_{n+n'+lN}^y
+\Delta S_n^z S_{n+n'+lN}^z\right) ,
\label{OriHam}
\end{eqnarray}
where we take the anisotropic parameter 
$\Delta =\frac{1}{2}g(g-1)$ with an even integer $g$.
Note that the summation over $l$ is introduced
in (\ref{OriHam}) to
treat the long-range $1/r^2$ interaction consistently in a 
ring geometry\cite{ISMD2}. 
If we use  periodic boundary conditions, this model 
reduces to the Haldane-Shastry model with the $\sin^{-2}$
exchange\cite{Hal,Sha}. We impose here twisted boundary conditions
instead,
\begin{equation}
S_{n+lN}^\pm=e^{\pm 2\pi il\phi}S_n^\pm ,\quad
S_{n+lN}^z=S_n^z .
\end{equation}
By the use of the gauge transformation 
$S_n^\pm = e^{\pm 2\pi in\phi /N}\widetilde S_n^\pm$
and $S_n^z=\widetilde S_n^z$,
we can treat the system
with periodic boundary conditions. 
In the following analysis, we restrict ourselves to 
rational values of the twist angle 
\begin{equation}
\phi =\frac{p}{q},\quad p\in{\bf Z},q\in{\bf N}.
\end{equation}
In this condition the Hamiltonian is well defined and
can be written down explicitly,
\begin{eqnarray}
H=\left(\frac{\pi}{N}\right)^2\sum_{n<n'}
\biggl[&&\frac{1}{2}\left(J_\phi (n-n')
\widetilde S_n^+\widetilde S_{n'}^-+h.c.\right)
\nonumber\\
&&+\Delta J_{0}(n-n')
\widetilde S_n^z\widetilde S_{n'}^z\biggr]
\label{Ham}
\end{eqnarray}
on a periodic ring with $N$ sites, $1\le n<n'\le N$, where
we have introduced the exchange coupling including 
the effect of $\phi$,
\begin{equation}
J_\phi (n)\equiv\frac{1}{q^2}\sum_{m=0}^{q-1}
e^{2\pi ip(n+mN)/qN}\sin^{-2}\left[\frac{\pi (n+mN)}{qN}\right].
\end{equation}
This model is hence regarded as a
``deformation'' of the periodic Haldane-Shastry model,
threaded by ``effective flux'' $\phi$.
The coupling 
 $J_\phi (n)$ has the following properties; (a)
$J_\phi (n+N)=J_\phi (n)$, (b) $J_\phi^*(n)=J_\phi (-n)$
and (c) $J_{0}(n)=\sin^{-2}(\pi n/N)$.
In the hard-core boson representation $\widetilde S_n^-=b_n^\dagger$ and 
$\widetilde S_n^z=1/2-b_n^\dagger b_n$, the Hamiltonian can be expressed by
$H=\left(\frac{\pi}{N}\right)^2(H_h+H_i+e_1+e_2)$, where
\begin{eqnarray}
H_h&&=\frac{1}{2}\sum_{n\ne n'}
J_{\phi}(n-n')b_{n'}^\dagger b_n ,\nonumber\\ 
H_i&&=\Delta\sum_{n<n'}J_{0}(n-n')b_{n'}^\dagger 
b_{n'}b_n^\dagger b_n ,
\label{HamBos}
\end{eqnarray}
$e_1=\frac{1}{24}\Delta N(N^2-1)$
and $e_2=-\frac{1}{6}\Delta M(N^2-1)$.
Here, $M$ denotes the number of bosons, i.e., the number of 
down-spins.


Motivated by Haldane's and Shastry's solution for 
$\phi =0$\cite{Hal,Sha}, 
we consider the wave function 
\begin{equation}
\widetilde\psi (\{n_i\})=\prod_jz^{Jn_j}\prod_{i<j}d(n_i-n_j)^g ,
\label{Jas}
\end{equation}
where $z\equiv\exp (2\pi i/N)$
and $d(n)=\sin(\pi n/N)$.
Note that the wave function in the original basis
can be obtained by a gauge transformation, 
$\psi (\{n_i\})=\prod_j z^{\phi n_j}\widetilde\psi (\{n_i\})$.
We thus find the quantum number $J$ is related to the 
crystal momentum $K(\phi)$ via 
$K(\phi )={2\pi M \over N}(J+\phi )$ modulo $2\pi$ .


We now wish to prove that (\ref{Jas}) is indeed an
eigenfunction of the Hamiltonian (\ref{HamBos}).
Similar calculations to Haldane\cite{Hal} show that
the kinetic term acts on (\ref{Jas}) as
\begin{equation}
\frac{H_h\widetilde\psi(\{n_i\})}{\widetilde\psi(\{n_i\})}
=\frac{1}{2}\sum_{n=1}^{N-1}J_\phi (n)z^{Jn}\sum_i\prod_{j(\ne i)}
(1-g_{ij}^{(n)})^{g/2},
\label{HopWav}
\end{equation}
where 
$g_{ij}^{(n)}\equiv ((1-z^n)z_i^2+(1-z^{-n})z_j^2)/(z_i-z_j)^2$
with $z_i\equiv z^{n_i}$.
By expanding the last factor in power series of 
$(1-z^n)^s(1-z^{-n})^t$ 
we can prove that r.h.s of eq.(\ref{HopWav}) 
consists of constants minus 
$H_i\widetilde\psi(\{n_i\})/\widetilde\psi(\{n_i\})$,
which should exactly cancel the interaction term.
A key quantity to prove this result is
\begin{equation}
S_{st}^\phi (J)=\frac{1}{4}
\sum_{n=1}^{N-1}J_\phi (n)z^{Jn}(1-z^n)^s(1-z^{-n})^t,
\label{KeyFor}
\end{equation}
where $J$ is an integer in the range 
$t-\phi\le J\le N-\phi -s$. One can easily obtain the following 
formulas: 
(a) $S_{st}^\phi (J)=0$ for $3\le s+t\le \min(J+\phi,N-J-\phi)$,
(b) $S_{st}^\phi (J)=(-)^s$ for $s+t=2$,
(c) $S_{st}^\phi (J)=(-)^s(J+\phi-N/2)-1/2$ for $s+t=1$ and
(d) $S_{00}^\phi (J)=\{[J+\phi]+(1-N)/2\}(J+\phi)-
[J+\phi]([J+\phi]+1)/2+(N^2-1)/12$. 
Here $[a]$ denotes the largest integer 
which does not exceed the value of $a$.
Taking into account these formulas,
we conclude that for $\frac{g}{2}(M-1)\le J+\phi\le N-\frac{g}{2}(M-1)$,
the wave function (\ref{Jas}) is an eigenstate with the eigenvalue
$E_{\rm t}(\phi )=\left(\frac{\pi}{N}\right)^2(e+E(\phi ))$, 
where
\begin{eqnarray}
e&=&\frac{1}{6}(N^2-1)\left\{
\frac{1}{4}N\Delta +M(1-\Delta )\right\},\nonumber\\
E(\phi )&=&\frac{1}{12}g^2M(M^2-1)  \nonumber\\
&&+M\bigl\{(2[J+\phi ]+1-N)(J+\phi ) \nonumber\\
&&~~~~~~~-[J+\phi ]([J+\phi ]+1)\bigr\}.
\label{ExaSpe}
\end{eqnarray}
This gives the exact energy for the twisted 
Haldane-Shastry model as a function of $\phi$.


Henceforth we concentrate on the isotropic model $\Delta =1~(g=2)$
with an even-integral $N$.
It is seen from (\ref{ExaSpe}) that the absolute 
ground state with the lowest energy is given by the 
quantum number $J=\frac{N}{2}$ for $ 0 \le \phi \le 1/2$,
$J=\frac{N}{2}-1$ for $ 0 \le \phi \le 1/2$, etc.
However, this may not be the case for the 
``adiabatic'' ground state which evolves  continuously 
from the ground state at $\phi =0$. 
The period of the spectral flow is determined by 
that for the adiabatic ground state. 
Let us now increase $\phi$ gradually from 0
and follow the adiabatic ground state specified by the same
quantum number at $\phi =0$. One can see from (\ref{ExaSpe}) that 
for $0\le\phi\le 1$, the adiabatic ground state  
continuously evolves with the fixed quantum 
number $J=\frac{N}{2}$,  making 
a level-crossing with the state specified $J=\frac{N}{2}-1$ 
at $\phi=1/2$. 
At $\phi =1$, the adiabatic ground state becomes 
the first excited state. 
At this point, as seen from (\ref{ExaSpe}), the spectral flow has a 
cusp structure. So, 
it is not straightforward to 
follow the ground state through this point. Nevertheless, 
adopting the idea of the Bethe ansatz
which will be introduced momentarily,
we can determine the natural spectral flow. We then find  
that the spectral flow is further traced by
the state of $J=\frac{N}{2}-2$\cite{Com3} beyond $\phi =1$ point, 
and finally returns to the absolute ground state 
at $\phi =2$.  
The spectral flow of the adiabatic ground state 
is thus obtained from (\ref{ExaSpe}),
\begin{equation}
E(\phi )=\left\{
\begin{array}{ll}\frac{N}{2}\phi
-\frac{1}{12}N(N^2+2),& \quad {\rm for}\quad 0\le\phi\le 1\\
-\frac{N}{2}\phi
-\frac{1}{12}N(N^2-10),& \quad {\rm for}\quad 1\le\phi\le 2
\end{array}\right.
\label{GroSpe}
\end{equation}
Therefore, one can see that 
the period of the ground state is 2 (in unit of $2\pi$),
which, as will be shown below,  directly reflects 
fractional exclusion statistics. To see 
our statement more clearly, we have 
carried out the numerical-diagonalization calculation for
finite-size systems. In Fig.1, we have drawn the full 
spectral flow calculated for the $N=6$ system.
An interesting point in this figure is that one cannot see
any level-repulsions among energy levels at $\phi=1$, 
and there are singular cusps there, which indeed accord 
with (\ref{ExaSpe}) and (\ref{GroSpe}). 
These characteristic features show a sharp contrast to  
ordinary quantum systems
for which level-repulsions due to irrelevant operators
make the spectral flow smooth at $\phi=1$ 
for a finite system\cite{SutSha}.
These remarkable facts reflect that the 
$1/r^2$ model is a kind of fixed-point Hamiltonian without
any irrelevant perturbations\cite{Hal_f}. Also, 
the linear dependence of the 
spectral flow as a function of $\phi$, which seems
somewhat peculiar, is related to the higher symmetry of
the model, and reflects a special dispersion relation
(see eq.(\ref{Dis})).  


We now give an asymptotic-Bethe-ansatz (ABA) description 
of the energy spectrum.
The advantage to use this method is that 
one can directly observe the 
fractional exclusion statistics in the spectral flow.
For this purpose, let us first derive the single-particle dispersion
relation.  In the momentum space, the hopping Hamiltonian is written as
$H_h=\sum_{k=1}^N\varepsilon (\widetilde k)b_k^\dagger b_k+\frac{M}{6}(N^2-1)$,
where $[b_k,b^\dagger_{k'}]=\delta_{k,k'}$ and 
$\widetilde k=k+\phi$ ($k$ is integral, so 
$\widetilde k$ is fractional in general).
The single-particle dispersion relation fulfills 
$\varepsilon (\widetilde k+N)=\varepsilon (\widetilde k)$ and
is given by
\begin{equation}
\varepsilon (\widetilde k)=\left(2[\widetilde k]
-N+1\right)\widetilde k
-[\widetilde k]\left([\widetilde k]+1\right) ,
\label{Dis}
\end{equation}
which holds for $0\le\widetilde k\le N$.
For other values of $\widetilde k$, one can use the periodicity of 
$\varepsilon (\widetilde k)$. When $\phi =0$, 
$\widetilde k$ becomes integral $k$, resulting in 
the well-known quadratic form $\varepsilon (k)=k(k-N)$
\cite{Hal,Sha}.
Note that such a quadratic dispersion is valid
only for integral $k$, and we have to use eq.(\ref{Dis})
for a fractional value of $ \widetilde k$.
Following the idea of ABA\cite{ISMD2}, 
the total Hamiltonian for the isotropic case ($\Delta =1$)
can be diagonalized as
$H=\left(\frac{\pi}{N}\right)^2(H_0+e)$, where
$H_0 =\sum_{k=1}^N\varepsilon (\widetilde k)b_k^\dagger b_k$
and $e=\frac{1}{24}N(N^2-1)$.
What is remarkable is that all the interaction effects are 
completely incorporated as statistical interactions, 
which can be explicitly formulated  by the 
following ABA equation for the rapidity
$\widetilde k_i$,
\begin{equation}
\widetilde k_i=I_i+\phi +\frac{1}{2}\sum_{j(\ne i)}{\rm sgn}
(\widetilde k_i-\widetilde k_j).
\label{ABA}
\end{equation}
This gives  the exact momentum and energy
in simple forms,
\begin{equation}
 K(\phi )=\frac{2\pi}{N}\sum_{i=1}^M\widetilde k_i ,
\hskip 5mm
E(\phi )=\sum_{i=1}^M\varepsilon (\widetilde k_i) .
\label{Ene}
\end{equation}
Note that the two-body phase-shifts in eq.(\ref{ABA}) 
represent the effects of statistical interactions.


Now let us study the ground state energy as a function of $\phi$
in terms of the ABA again.
Here the ground state means the adiabatically followed ground state.
Note first that  the ground state at $\phi =0$ is given by 
$\widetilde k_i=k_{i}=1,3,...,N-1$, corresponding
to the set of consecutive quantum numbers 
$I_i =\left(\frac{N}{2}-1\right)/2+i$ with $i=1,2,\cdots,N/2$.
This state can be clearly specified by using the idea of {\it motif}
in fractional exclusion statistics,
in which we take into account the repulsion effect due to 
the phase shift in (\ref{ABA}) by substituting  0's uniformly among 
a sequence of consecutive 1's (occupied momenta)
\cite{HHTBP}. For example, the 
ground state for free fermions is labeled  by
the motif $000011\cdots110000$, reflecting the Pauli
principle. For the Haldane-Shastry model, the phase shift 
in eq.(\ref{ABA}) is given by 
${1 \over 2}(g-1)\sum_{j(\ne i)}{\rm sgn}(k_i-k_j)$
with the statistical interaction $g=2$.
So, the ground state configuration in the momentum 
space may be $g$-times enlarged  with respect to 
free fermions, which is denoted by the 
motif $0101\cdots1010$ for $g=2$. Switching on the twist, 
we have from (\ref{ABA}),
$\widetilde k_i=k_{i}+\phi$ with $i=1,2,\cdots ,\frac{N}{2}$.
Namely, $\phi$ merely shifts the momentum $k_i$ uniformly,
and does not change the set of the quantum numbers $\{I_i\}$
of the ground state.
By substituting the above momenta into eq.(\ref{Ene}),
we can reproduce the  exact spectral flow of
eq.(\ref{GroSpe}). An important point in this calculation
with the ABA is that one can clearly see a 
physical meaning of the period 2 in terms of 
exclusion statistics.
Namely the above spectral flow 
is correctly interpreted  by the motif 
constructed by $\widetilde k_i$,
\begin{equation}
010101\cdots 101010\stackrel{\displaystyle{\rightarrow}}{\leftarrow} 
001010\cdots 010101 ,
\end{equation}
where the arrow means that we add a unit 
flux $\delta\phi =1$ in this process.
We emphasize here that the motif picture is valid not only for the 
integral $\phi$ but also for every fractional $\phi$.
It is essentially classified by the motif at $\phi =0$.
Through the above analysis using the motif for the ABA,
we can clearly see that {\it the period $2$ observed in the 
Haldane-Shastry model directly reflects the fractional exclusion
statistics with the statistical interaction $g=2$}.
Also for excited states,  we can reproduce the  correct 
spectral flow in terms of the motif.
For more general cases with even integers $g$,
the period of the spectral flow can be regarded as $g$ \cite{FukKaw}.

We recall here that the period 2 was observed previously
in the ordinary Heisenberg model with nearest-neighbor 
interaction\cite{SutSha}. It should be noticed,  
 however, that its origin is quite different 
from the present one. In fact one may not observe the 
properties of fractional exclusion statistics 
in the spectral flow  for the nearest-neighbor model,
because irrelevant perturbations among spinon excitations 
actually determine the period 2 in that case
\cite{SutSha,FukKaw},
in contrast to the Haldane-Shastry model without
such irrelevant perturbations. This implies that 
the Haldane-Shastry model with $1/r^2$ interaction is 
an idealized model for which one can see 
exclusion statistics clearly in the spectral flow.
It should be pointed out here that ``ideal particles" obeying 
fractional exclusion statistics indeed appear in the calculation of 
the exact dynamical correlation functions for 
$1/r^2$  systems\cite{DynCor}.

Finally we wish to briefly mention 
the symmetry property of our system. 
We naively expect that SU(2) symmetry is reduced
to U(1) by applying the magnetic flux.
We recall here, however, that the Haldane-Shastry
model has Yangian symmetry larger than SU(2)
at $\phi=0$\cite{HHTBP}. We will show that a part 
of Yangian symmetry still survives even for 
a finite $\phi$.
The generators for level-1 SU(2) 
Yangian  are defined by\cite{HHTBP}
\begin{equation}
Q^a=\frac{1}{2}\sum_{n=1}^N\sum_{n'=1}^{N-1}\sum_{l=-\infty}^\infty
\frac{1}{n'+lN}\epsilon^{abc}S_n^bS_{n+n'+lN}^c,
\end{equation} 
where $a=1,2,3$.
We must take the summation with respect to $l$ 
so as to make it converge.
The third component is explicitly given by 
\begin{equation}
Q^3=\frac{1}{2i}\sum_{n<n'}\left(
W_\phi (n-n')\widetilde S_n^+\widetilde S_{n'}^-
-W_\phi^*(n-n')\widetilde S_n^-\widetilde S_{n'}^+\right) ,
\label{YanCar}
\end{equation}
where
\begin{equation}
W_\phi(n)=\frac{\pi}{qN}\sum_{m=0}^{q-1}e^{2\pi ip(n+mN)/qN}
\cot\left[ \frac{\pi (n+mN)}{qN}\right] .
\end{equation}
This function fulfills 
$W_\phi(n+N)=W_\phi(n)$ and $W_\phi(-n)=-W_\phi^*(n)$.
A tedious but straightforward 
calculation proves that  
$Q^3$ still commutes with  $H$; $[H, Q^3]=0$
for the isotropic case.
Therefore, we can see  that 
a nontrivial conserved quantity
associated with the SU(2) Yangian symmetry still persists
even for the finite flux, which may have a deep
relation to characteristic properties observed in 
the spectral flow. 

In this paper, we have been concerned with 
rational values of the twist angle $\phi$. We think that 
the analysis similar to the present one could be also applied 
to irrational cases to reveal the nature of fractional 
statistics, though we have not yet succeeded to obtain the 
exact solution to the irrational cases.

The authors would like to thank  Y. Kuramoto, Y. Kato, S. Fujimoto,
T. Yamamoto and H. Awata for valuable discussions.
This work is supported by the Grant-in-Aid from the Ministry of
Education, Science and Culture, Japan.



\begin{figure}[h]
\epsfxsize=7cm 
\vspace{0.5cm}
\caption{Exact spectral flow for the 
$N=6$ system with $\Delta=1$. 
In order to change the value of $\phi =p/q$,
integer $p$ is varied consecutively with $q=100$ being fixed.}
\end{figure}


\begin{references}
\bibitem{Com3} Note that at $\phi =1$
two states with $J=N/2-2$ and $N/2$
have the same crystal momentum.
\end{references}
\end{document}